\documentclass[prb,aps,showpacs,twocolumn,preprintnumbers,
amsmath,amssymb,superscriptaddress,longbibliography]{revtex4-2}
\usepackage[english]{babel}
\usepackage{amsmath,amssymb,amsfonts}
\usepackage{graphicx}
\usepackage[colorlinks=True,linkcolor=blue,citecolor=blue,urlcolor=blue]{hyperref}

\usepackage{booktabs}
\usepackage[dvipsnames]{xcolor}
\usepackage{comment}
\usepackage{braket}
\usepackage{bm}
\usepackage{bbm}
\usepackage{braket}
\usepackage{float}
\usepackage{multirow}
\usepackage{longtable}
\usepackage[normalem]{ulem}
\usepackage{array}
\usepackage{makecell}
\usepackage{subfigure}
\usepackage{booktabs}
\usepackage{multirow}
\usepackage{graphicx}
\graphicspath{{NewFigures/}}
\usepackage{dcolumn}
\usepackage{bm}
\usepackage{longtable}
\usepackage[dvipsnames]{xcolor}

\newcolumntype{C}{>{$}c<{$}}

\allowdisplaybreaks
\DeclareMathAlphabet{\zc}{OT1}{pzc}{m}{it}

\def\ket#1{|#1\rangle }
\def\bra#1{\langle #1 |}

\begin{document}
%

\title{Non-Hermitian ultra-strong clustering through interaction-induced caging}

\author{Mengjie Yang}
\affiliation{State Key Laboratory of Advanced Optical
Communication Systems and Networks, School of Physics and Astronomy, Shanghai Jiao Tong University, Shanghai 200240, China}
\affiliation{Department of Physics, National University of Singapore, Singapore 117551, Singapore}

\author{Luqi Yuan}
\email{yuanluqi@sjtu.edu.cn}
\affiliation{State Key Laboratory of Advanced Optical
Communication Systems and Networks, School of Physics and Astronomy, Shanghai Jiao Tong University, Shanghai 200240, China}

\author{Ching Hua Lee}
\email{phylch@nus.edu.sg}
\affiliation{Department of Physics, National University of Singapore, Singapore 117551, Singapore}

\date{\today}

\begin{abstract}
We uncover a new mechanism whereby the triple interplay of non-Hermitian pumping, bosonic interactions and nontrivial band topology leads to strong boson clustering. The extent of boson clustering goes beyond what is naively expected from the interaction-induced trapping of non-Hermitian pumped states and is based on an emergent caging mechanism that topological boundary modes can further enhance. Beyond our minimal model with two bosons, this caging remains applicable for generic many-boson systems subject to a broad range of density interactions and non-Hermitian hopping asymmetry. Our novel mechanism for particle clustering would inspire fundamental shifts in our comprehension of many-body non-Hermitian dynamics and open new avenues for controlling and manipulating bosons. 
\end{abstract}

\maketitle


\section{Introduction}

There has been growing excitement in interplay between many-body interactions and the non-Hermitian skin effect (NHSE)~\cite{mu2020emergent,li2021impurity,yang2021exceptional,kawabata2022many,alsallom2022fate,shen2023observation,ozturk2021observation,gong2022anomalous,zhang2022symmetry,roccati2022exotic,shen2022non,gliozzi2024many,kim2024collective,shimomura2024general,yoshida2024non,roccati2024hermitian}. This stems from the non-locality and sensitivity of the NHSE~\cite{li2021impurity,shen2022non}, which profoundly impacts particle localization~\cite{yang2021exceptional,kawabata2022many,alsallom2022fate,shen2023observation} and localization dynamics~\cite{ozturk2021observation,gong2022anomalous,roccati2022exotic}, which leads to unexpected consequences in the presence of quantum many-body interactions. These include the emergence of a many-fermion Fermi skin~\cite{mu2020emergent,shen2023observation}, exotic spin liquids from environmental couplings~\cite{yang2021exceptional}, spectral symmetry breaking~\cite{zhang2022symmetry}, unconventional quantum dynamics in non-Hermitian baths~\cite{gong2022anomalous,roccati2022exotic,roccati2024hermitian} and the non-Hermitian Mott skin effect~\cite{yoshida2024non}.

In this work, we uncovered a new mechanism whereby bosons can be made to condense very strongly through the triple interplay of non-Hermitian pumping, density interactions and non-trivial band topology. Due to the emergent non-locality of both the NHSE~\cite{li2021impurity,shen2022non} and the density interactions, the observed phenomenon contrasts starkly with the conventional behavior of interacting bosons, such as photons, which repel each other when they interact i.e. photon blockade~\cite{birnbaum2005photon, rabl2011photon, ridolfo2012photon, huang2018nonreciprocal}. In particular, we show that ultra-strong clustering, and not just boundary state accumulation, arises from an emergent caging mechanism that requires both non-Hermitian pumping and density interactions.

\section{Methods}

To demonstrate how non-Hermitian pumping, topology and bosonic interactions can interplay to cause unexpectedly strong bosonic clustering, we consider a minimal interacting 1D bosonic lattice model that contains topological edge modes at its boundary sites:
\begin{equation}
\begin{aligned}
    H=&\sum_{x=1}^L t_L b_{2x-1}^{\dagger} b_{2x}+t_R b_{2x}^{\dagger} b_{2x-1}\\&+\sum_{x=1}^{L-1}t_0\left(b_{2x}^{\dagger} b_{2x+1}+b_{2x+1}^{\dagger} b_{2x}\right)+\frac{U}{2} n_{x_0}^2, 
\end{aligned} \label{eq:H}
\end{equation}
which is a Su-Schrieffer-Heeger (SSH) lattice~\cite{su1979solitons} with asymmetric intercell ($t_L$ and $t_R$) hoppings and unit intracell hoppings, equipped with a density interaction $\frac1{2}U n_{x_0}^2$ at site $x_0$, $U\geq 0$. Here $b_x$ ($b_x^\dagger$) is the bosonic annihilation (creation) operator at sites $x=1,2,\ldots, 2L$, with $n_{x_0}=b_{x_0}^\dagger b_{x_0}$ the boson number operator. Non-Hermitian pumping~\cite{lee2016anomalous,alvarez2018non,yao2018edge,kunst2018biorthogonal,lee2019anatomy,zhang2021observation,yang2022concentrated,li2022non,shen2022non,lin2023topological,yang2024percolation} occurs towards the left if $r=t_L/t_R> 1$, accumulating the bosons towards the left boundary. However, when multiple indistinguishable bosons exist, this accumulation would also be affected by the repulsive ($U>0$) interaction at site $x_0$, with strength depending on bosonic occupancy. In this work, we have chosen the interaction $\frac1{2} U n_{x_0}^2$ that mimics the simplest possible nonlinearity in the mean-field limit, different from the usual Bose-Hubbard interaction~\cite{jaksch1998cold,birnbaum2005photon,winkler2006repulsively,rabl2011photon,ridolfo2012photon} by a local density shift~\cite{chang2014quantum,xiao2022bound,yuan2020creating,cheng2021arbitrary,barbiero2020bose}.

\begin{figure*}
    \includegraphics[width=0.96\textwidth]{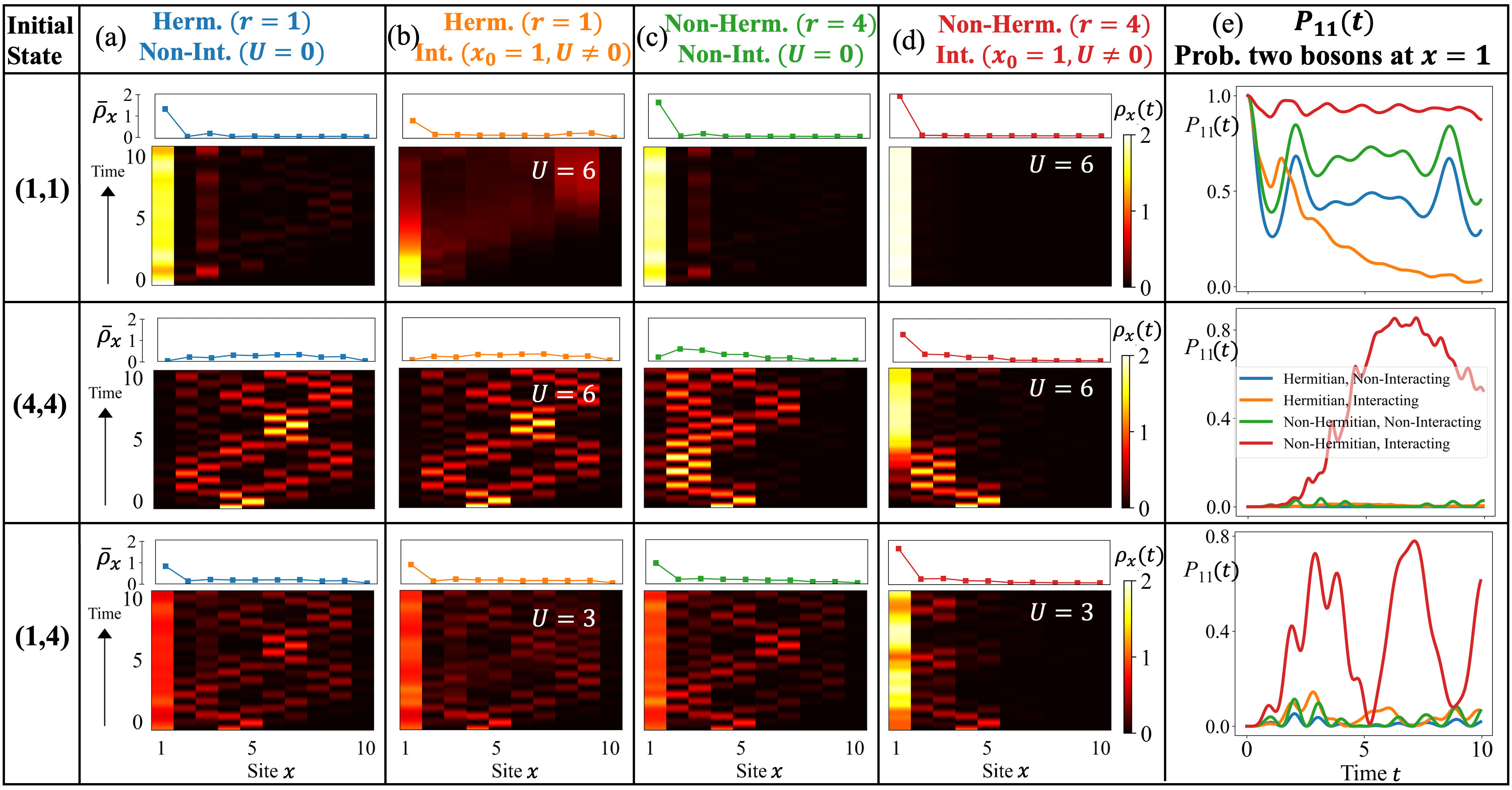}
    \caption{Two-boson correlation dynamics and associated spatial density distributions in a boundary-interacting system. The dramatic enhancement of bosonic clustering at site $x=1$ is captured by $P_{11}$ [Eq.~\ref{eq:Pmn}], which measures the probability of finding both bosons at this site. This correlation is significantly stronger (red curves in panel e) when both the non-Hermitian skin effect (NHSE, $r>0$) and boundary interaction ($U>0$) are present, demonstrating their synergistic effect on boson trapping. This enhanced clustering is studied for different initial states $(1,1)$, $(4,4)$, and $(1,4)$ evolving under $H$ [Eq.~\ref{eq:H}]. Supporting this observation, panels (a-d) show the spatial density evolution $\rho_x(t)$ [Eq.~\ref{eq:bosonnum}] over time $t \in [0, 10]$, with corresponding time-averaged profiles $\bar{\rho}_i$ [Eq.~\ref{eq:timeaverage_rho}] plotted above. The density distributions contrast four scenarios: Hermitian ($r=1$) without (a) and with (b) interaction, and non-Hermitian ($r=4$) without (c) and with (d) interaction. Notably, the interaction term $U$, despite being typically repulsive, acts as an effective trap at $x_0=1$ during non-equilibrium evolution when combined with leftward NHSE, as evidenced by the bright regions in the heatmaps. Parameters: $t_0=3$, and for $r=4$, $t_L=1.6$ and $t_R=0.4$.}
    \label{fig:mainresult}
\end{figure*}

While one might naively expect a repulsive ($U>0$) density interaction to primarily suppress the NHSE, the observed behavior can be dramatically different even with just two bosons. Two distinct measures of particle accumulation can be defined: the spatial density accumulation and the many-body correlation. An initial 2-boson state $\ket{\phi(t=0)}$ evolves according to the Schrodinger equation $\ket{\dot{\phi}(t)}=-iH\ket{\phi(t)}$, and can be expressed as
\begin{equation}
    \ket{\phi(t)}=\frac{1}{\sqrt{2}}\sum_{x=1}^{2L}\sum_{x'=1}^{2L}v_{xx'}(t)\ket{(x,x')},\label{eq:twobosonstate}
\end{equation}
where $v_{xx'}(t)=v_{x'x}(t)$ is the amplitude of the basis state $\ket{(x,x')}=b_x^\dagger b_{x'}^\dagger \ket{0}$ containing indistinguishable bosons at sites $x,x'$. 

We define the 2-boson density at site $x=1,2,\ldots,2L$ by
\begin{equation}
    \rho_x(t)=\frac{\bra{\phi(t)}b_x^\dagger b_x\ket{\phi(t)}}{\langle\phi(t)|\phi(t)\rangle}, 
    \label{eq:bosonnum}
\end{equation}
with $0 \leq \rho_x(t)\leq 2$. Associated with it is the time-averaged spatial density
\begin{equation}
    \bar{\rho}_x=\frac{1}{T}\int_{0}^{T}\rho_x(t') dt',
    \label{eq:timeaverage_rho}
\end{equation}
where $T$ is the simulation duration.

While $\rho_x(t)$ and $\bar \rho_x$ reveal where the bosons localize on the lattice, a high value of $\rho_x(t)$ or $\bar \rho_x$ can physically arise either due to strong single-boson localization at site $x$, or moderate double-boson clustering at the same site. To quantify this important distinction, we also examine the two-boson correlation probability
\begin{equation}
P_{xx'}(t) = 
\begin{cases} 
|v_{xx'}(t)|^2 & \text{if } x = x', \\
|\sqrt{2}v_{xx'}(t)|^2  & \text{if } x \neq x',
\end{cases}\label{eq:Pmn}
\end{equation}
which represents the probability of observing a boson at sites $x$ and $x'$ at the same time $t$. To reveal underlying trends in the evolution of the two-boson correlation, we also define the time-smoothed correlation
\begin{equation}
\bar{P}_{xx'}(t) = \frac{1}{\Delta t} \int_{t-\Delta t}^{t} P_{xx'}(t') dt',
\label{eq:smoothedPmn}
\end{equation} 
which removes temporal oscillations shorter than a prescribed timescale $\Delta t$.

\section{Results}
\subsection{Ultra-strong non-Hermitian Bosonic clustering}

We first consider scenarios where the density interaction is at the leftmost (boundary) site $x_0=1$. Figs.~\ref{fig:mainresult}(a-d) showcase the evolution of the dynamical density $\rho_x(t)$ [Eq.~\ref{eq:bosonnum}] and its time-average $\bar \rho_x$ [Eq.~\ref{eq:timeaverage_rho}] of three illustrative initial 2-boson state configurations.

For the initial state $(1,1)$ (Top Row) where both bosons are already on $x_0=1$, the density interaction indeed repels the bosons in the Hermitian limit, as evidenced in the suppressed $\bar \rho_x$ density at $x=x_0=1$ across the Hermitian interacting vs. non-interacting cases (orange vs. blue). Somewhat expected, this suppression vanishes when the NHSE counteracts the repulsion by pumping the bosons towards $x_0=1$ (green and red). However, for the initial state $(4,4)$ (Center Row) where both bosons are initially far from $x_0=1$, the boson density oscillates and spreads out, failing to accumulate appreciably at $x_0=1$ except when both the non-Hermiticity and interaction are present (red). Indeed, in this non-equilibrium scenario, the $U>0$ density term behaves more like a local potential well that traps the $\rho_x(t)$ at $x=x_0$, rather than repulsion. Qualitatively similar behavior is observed in the density evolution for initial state $(1,4)$ (Bottom Row), when one of the bosons is initially already at $x_0=1$ and is as such unaffected by the NHSE.

What is striking, however, is the ultra-strong two-boson clustering at boundary site $x_0=1$ when non-Hermiticity ($r>1$) and the density interaction ($U>0$) are both present. This is revealed in the dynamical $P_{11}(t)$ [Eq.~\ref{eq:Pmn}] plots in Fig.~\ref{fig:mainresult}(e), which shows the probability of both bosons condensing at $x=1$. From the Top Row, when both bosons are prepared at $x=1$ (initial state =(1,1)), they can only remain there if the interaction and non-Hermiticity are simultaneously present (red). But most surprising is what happens when at least one boson is initially away from $x=1$ (Center and Bottom Rows): we observe \emph{overwhelmingly} higher $P_{11}$ only when non-Hermiticity and interaction simultaneously interplay (red), compared to having either on their own (yellow, green). This suggests that trapping both bosons and suppressing photon blockade~\cite{rabl2011photon,ridolfo2012photon,birnbaum2005photon} hinges on an emergent consequence of this interplay, a fact not evident from the density evolution alone [Fig.~\ref{fig:mainresult}(a-d)].

\begin{figure}
    \centering
    \includegraphics[width=0.48\textwidth]{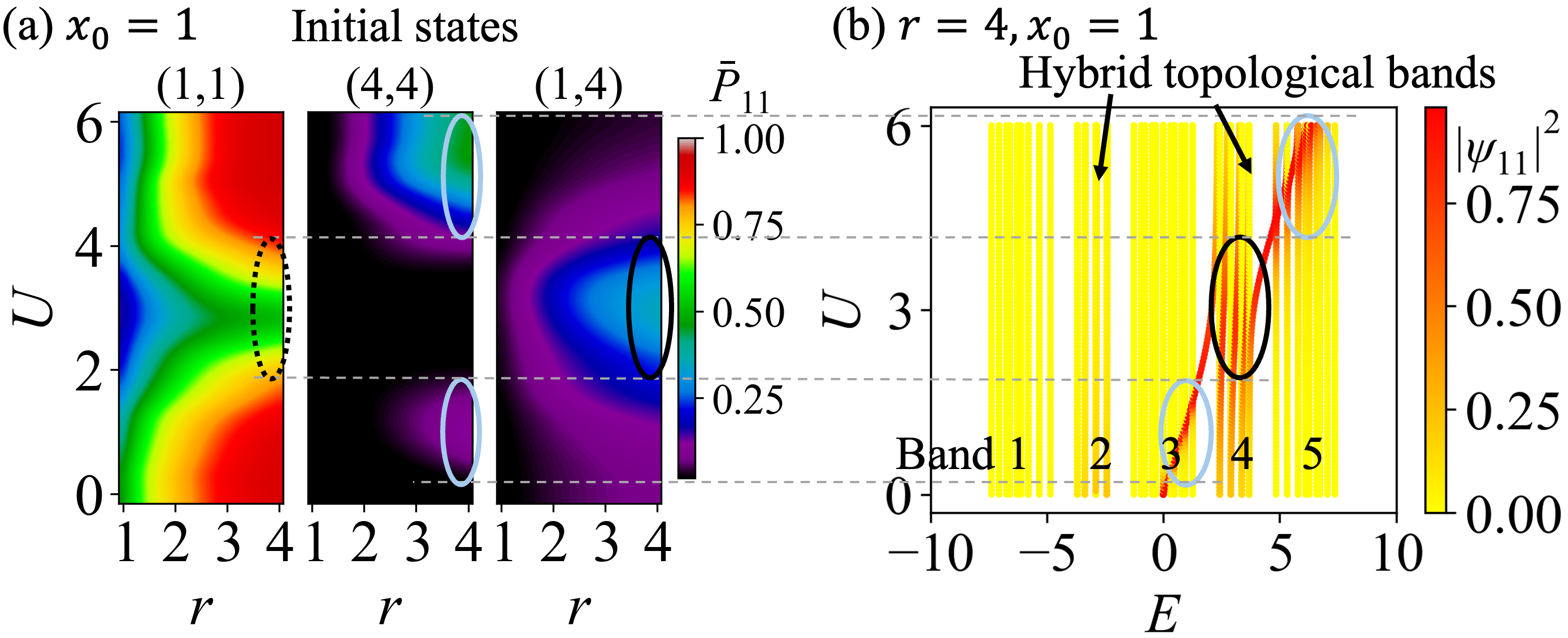}
    \caption{Extent of two-boson clustering due to a boundary density interaction at $x_0=1$ [Eq.~\ref{eq:H}], and its correspondence with the boundary clustering of 2-boson spectral bands. (a) Time-averaged two-boson clustering probability $\bar{P}_{11}$ [Eq.~\ref{eq:smoothedPmn}] at site $x=1$, in the parameter space of non-Hermitian hopping asymmetry $r$ and density interaction strength $U$. (b) Corresponding two-boson spectra (purely real) at $r=4$ ($t_L=1.6$ and $t_R=0.4), t_0=3$, which features five bands, with bands 2 and 4 being containing a pair of interaction-hybridized bulk and topological bosons. As $U$ increases, it creates a group of eigenstates at $E\approx U$ that exhibits strong clustering $\psi_{11}$ (red) at $x=x_0=1$. Strong $\psi_{11}$ clustering in the hybrid topological band 4 leads to suppressed $\bar P_{11}$ for the boundary-localized initial state $(1,1)$ (circled in dashed black), but enhanced $\bar P_{11}$ clustering for the initial state $(1,4)$ (solid black). By contrast, strong $\psi_{11}$ clustering in the bulk bands 3 and 5 (circled in light blue) corresponds to enhanced $\bar P_{11}$ for the bulk initial state $(4,4)$.}
    \label{fig:heatmap_spectra}
\end{figure}

To quantitatively characterize this two-boson clustering, i.e., $P_{11}(t)$ , we plot in Fig.~\ref{fig:heatmap_spectra}(a) the time-averaged clustering probability $\bar{P}_{11}=\bar{P}_{11}(T)$ with $\Delta t=T$ [Eq.~\ref{eq:smoothedPmn}], which is the correlation $P_{11}$ smoothed over the entire simulation duration $T$. The heatmap of $\bar P_{11}$ in the interaction strength vs. non-Hermiticity $U$-$r$ plane [Fig.~\ref{fig:heatmap_spectra}(a)] reveals significantly enhanced $\bar{P}_{11}$ (colored) for certain $(U,r)$ regions, compared to the $r=1$ Hermitian limit where $\bar{P}_{11}$ essentially vanishes (black). While $\bar{P}_{11}$ generally increases with $r$, optimal clustering $\bar{P}_{11}$ occurs at windows of $U$ that are highly dependent on the initial state, as will be explained below.

\subsection{Topological origin of ultra-strong bosonic clustering}

Intriguingly, for a density interaction at the $x_0=1$ boundary, it turns out that the ultra-strong boundary bosonic clustering is not just due to the interaction-NHSE interplay, but also relies crucially on SSH topology. This is evident from the 2-boson band structure plot in Fig.~\ref{fig:heatmap_spectra}(b), where each eigenenergy $E$ is colored according to the overlap $\psi_{11}=\langle(1,1)|\psi\rangle$ of its corresponding eigenstate $\ket{\psi}$ with $\ket{(1,1)}=(b_1^\dagger)^2\ket{0}$. Computed for our Hamiltonian $H$ [Eq.~\ref{eq:H}] at strong non-Hermitian hopping asymmetry $r=4$, it features 5 bands, with bands 2 and 4 from the hybridization of bulk and topological bands. This follows from the single-boson band structure~\cite{su1979solitons,stegmaier2021topological} with two symmetrically gapped bulk bands separated by in-gap topological zero modes (see Appendix.~\ref{app:spectra} for more details). The effect of the $U$ density interaction is to induce high $\psi_{11}$ overlap (red) successively from bands 3 to 5, as $U$ is increased. It is noteworthy that the robust dynamical trends persist even in systems with such entirely real energy spectra, indicating deeper underlying physical principles. In contrast, dynamics in systems with complex energies (Im$E>0$) are more straightforward, being dominated by states with the largest imaginary energy components.

To illustrate, with the initial state (1,1), suppressed $\bar P_{11}$at $U\approx 3$ corresponds to high $\psi_{11}$ overlap with band 4 [circled in black in Figs.~\ref{fig:heatmap_spectra}(a,b)]. This is because band 4 contains one bulk and one topological boson, which is not consistent with maintaining both bosons at $x_0=1$. As such, the same interaction strength of $U\approx 3$ leads to enhanced $\bar P_{11}$ (dashed black) for the initial state (1,4), where only one boson overlaps with the topological boundary mode. Likewise, $U\approx 1$ or $U\approx 5$ favors the $\bar P_{11}$ bosonic clustering for the bulk initial state (4,4), as circled in light blue, since that is when the $\psi_{11}$ overlap with the bulk bands (3 and 5) is strongest.

\subsection{Two-boson interactions as caging mechanisms}

\begin{figure}
  \includegraphics[width=0.48\textwidth]{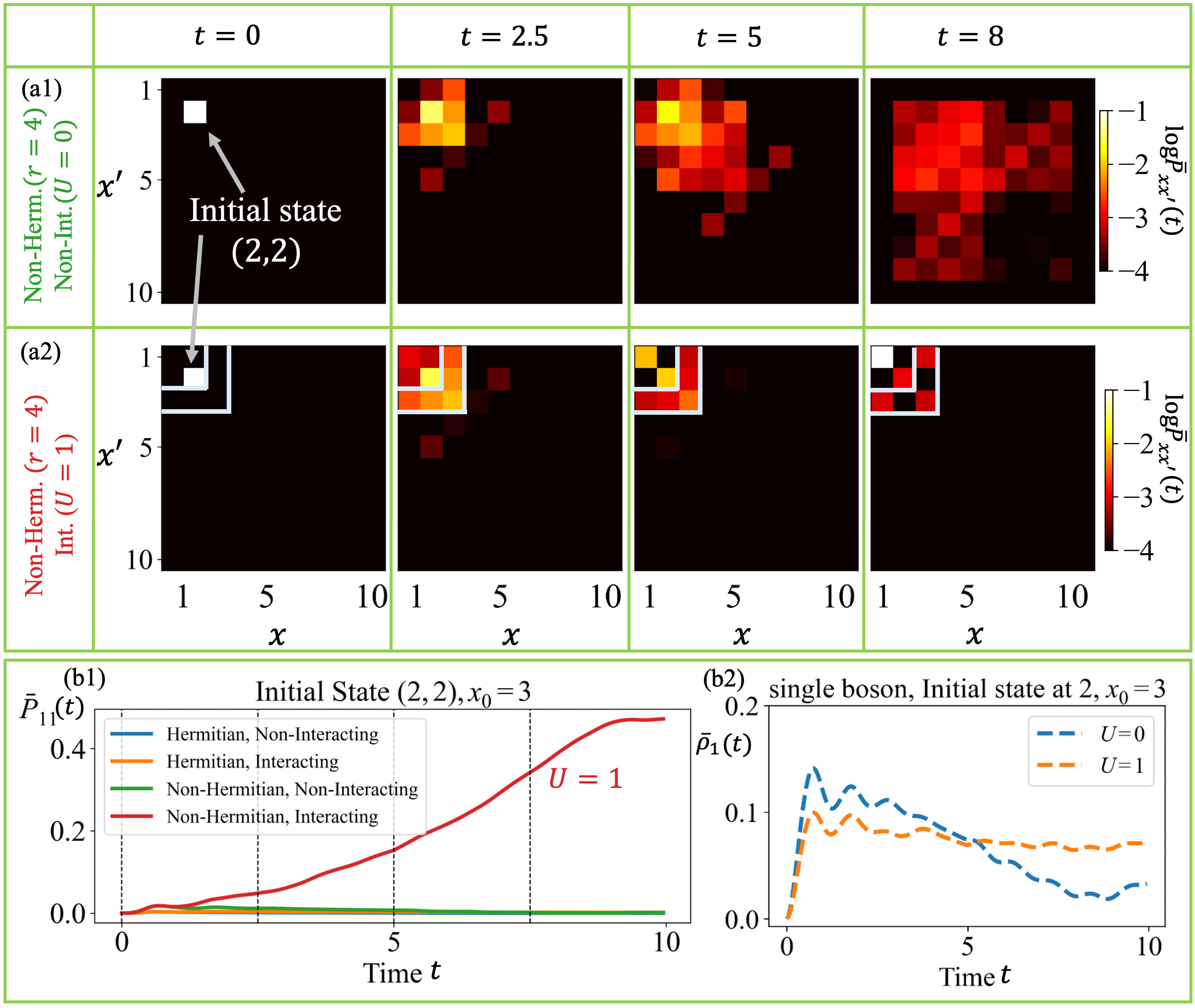}
  \caption{Ultrastrong boundary clustering $\bar P_{11}$ from interaction-induced caging, for initial state (2,2) within a $x_0=3$ cage for a 5 unit cell chain. (a) Snapshots of the time-smoothed ($\Delta t=2$) two-boson correlation probability $\bar{P}_{11}$ [Eq.~\ref{eq:smoothedPmn}] for (a1) non-interacting and (a2) interacting cases. The L-shaped cage in (a2) traps the bosons tightly. (b1) Evolution of probability of both bosons at the left boundary, with greatly enhanced $\bar P_{11}(t)$ only for the non-Hermitian interacting case (red). (b2) When only one boson is present, the boundary density $\bar\rho_1(t)$ remains low regardless of whether a $U$ barrier is present, showcasing that the caging mechanism is an interaction effect.}
  \label{fig:initial22x03}
  \end{figure}

    \begin{figure*}
        \includegraphics[width=0.98\textwidth]{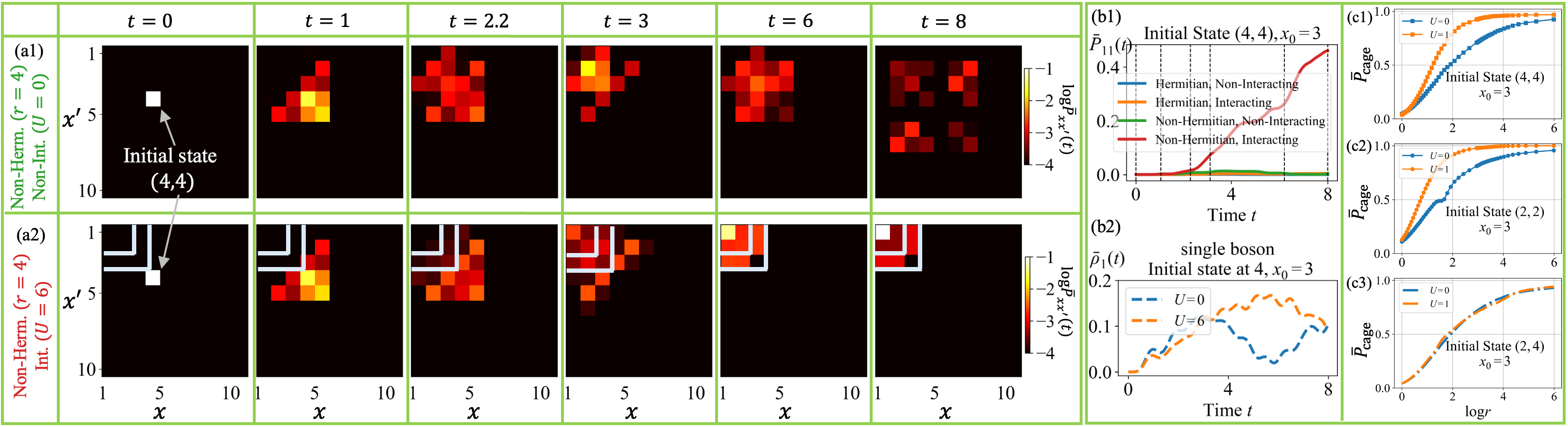}
        \caption{Persistence of ultrastrong boundary clustering $\bar P_{11}$ from interaction-induced caging, even for initial state (4,4) outside a $x_0=3$ cage.  (a) Snapshots of the time-smoothed ($\Delta t=2$) two-boson correlation probability $\bar{P}_{11}$ (Eq.~\ref{eq:smoothedPmn}) for (a1) non-interacting and (a2) interacting cases. The bosons still gradually penetrate the L-shaped cage due to the NHSE towards (1,1), and remain trapped in it after that ($t=3,8$).				
(b1) Eventually, $\bar P_{11}(t)$ is ultra-enhanced only for the non-Hermitian interacting case (red),  with (b2) negligible enhancement of boundary localization by $U$ in the single-boson case, similar to (b1,b2) of Fig.~\ref{fig:initial22x03}.
(c) Time-averaged two-boson probability $\bar P_\text{cage}$ within the $x_0\times x_0$ cage as a function of hopping asymmetry $r$. For initial states $(2,2)$ and $(4,4)$ with coincident bosons, the $U=1$ cases exhibit much larger $\bar P_\text{cage}$ due to interaction-induced caging. But for $(2,4)$,  non-coincident initial bosons do not interact appreciably and the effect of interactions is negligible.} 
       \label{fig:initial44x03}
        \end{figure*}
Interestingly, our density interaction $U$ can also act as a non-local cage when it is situated away from the boundary site $x=1$. In Fig.~\ref{fig:initial22x03}, we demonstrate how this interaction largely confines the evolved state to the left of $x_0$ i.e. $x\leq x_0$, if the initial state has both bosons in the same region. And, more elaborately, for an initial state with both bosons lying to the right of $x_0$, Fig.~\ref{fig:initial44x03} shows how the interaction will first act as a barrier to the evolved state, but then eventually still trap the particle flux that leaked through it due to NHSE pumping. Overall, this caging mechanism hence further enhances the ultra-strong boson clustering at the boundary. We shall fix the density interaction to be at $x_0=3$, and consider initial states $(2,2)$ and $(4,4)$ in Fig.~\ref{fig:initial22x03} and Fig.~\ref{fig:initial44x03} respectively.

This caging mechanism is most intuitively represented in the 2D configuration space $(x,x')$, where $x$ and $x'$ represent the positions of the two bosons\footnote{Here $(x,x')$ and $(x',x)$ refer to the same state due to bosonic statistics~\cite{winkler2006repulsively,valiente2008two,di2016two,gorlach2017topological,cheng2021arbitrary}.}. A key observation is that the nonlinear term $\frac1{2}Un_{x_0}^2$ can be interpreted as a non-local L-shaped ``cage'' [Figs.~\ref{fig:initial22x03}(a) and \ref{fig:initial44x03}(a)], with potential ``walls''~\cite{winkler2006repulsively,valiente2008two,di2016two,gorlach2017topological,cheng2021arbitrary} of height $\frac1{2}U$ across the entire lines $(x,x_0)$ and $(x_0,x)$ where only one boson is at $x_0$. These walls cross at $(x_0,x_0)$, where double bosonic occupancy at $x_0$ gives an even higher potential barrier of $2U$.

Fig.~\ref{fig:initial22x03}(a) compares the dynamical evolution of an initial state $(2,2)$ in the (a1) absence and (a2) presence of the density interaction $U$. In Fig.~\ref{fig:initial22x03}(a1) with $U=0$, the smoothed probability cloud $\bar P_{xx'}(t)$ [Eq.~\ref{eq:smoothedPmn}] of finding the bosons at $(x,x')$ spreads out freely and reflects against the boundaries, even though it is slightly amplified towards the $(1,1)$ corner due to leftwards NHSE pumping. However, in Fig.~\ref{fig:initial22x03}(a2), this spreading is drastically contained within the $3\times 3$ L-shaped cage formed by nonzero $U$ at $x_0=3$ [see Supplemental videos]. This caging, which requires both the $U$ interaction and directed amplification towards $(1,1)$, indeed leads to far enhanced boundary boson clustering $\bar P_{11}(t)$, as shown in the red curve in Fig.~\ref{fig:initial22x03}(b1). 

Interestingly, the caging mechanism protects the $\bar P_{11}$ clustering even when the initial state is outside the cage. In Fig.~\ref{fig:initial44x03}(a) with both bosons initially to the right of $x_0=3$ at $(4,4)$, their flux $\bar P_{xx'}(t)$ still gradually enters the $3\times 3$ cage due to the NHSE pumping towards $(1,1)$. Comparing the non-interacting (a1) with the interacting (a2) cases between $t=0$ to $2.2$, the interaction-induced potential walls only slow the diffusion into the cage slightly. However, once the bosons have entered the cage, they are subject to the same trapping mechanism as in Fig.~\ref{fig:initial22x03}, thereby also experiencing eventual ultra-strong clustering at $x=1$ [red in Fig.~\ref{fig:initial44x03}(b1)].

Notably, this caging mechanism is an emergent consequence of few-body density interactions, and cannot be replicated by an effective potential barrier in the single-particle context. Plotted in Figs.~\ref{fig:initial22x03}(b2) and \ref{fig:initial44x03}(b2) are the evolutions of the densities $\bar \rho_{x=1}$ of a \emph{single} boson in the same 1D SSH chain of Eq.~\ref{eq:H}, but with an on-site potential $U$ at $x_0=3$. The $U\neq 0$ cases (orange) do not trap the boson at $x=1$ boundary any more than the $U=0$ cases (blue), with the boundary boson density remaining far lower than the $\bar P_{11}(t)$ of the interacting NHSE 2-boson cases [red in Figs.~\ref{fig:initial22x03}(b1) and \ref{fig:initial44x03}(b1)].

The efficacy of the caging mechanism is further corroborated by plots of $\bar P_\text{cage}=\sum_{x,x'=1}^{x_0}\bar{P}_{xx'}$, the time-averaged two-boson probability within the cage $x,x'\leq x_0$. As evident in Figs.~\ref{fig:initial44x03}(c1-c3), $\bar P_\text{cage}$ increases steadily with the NHSE strength $r$, testimony to the crucial role of non-Hermitian pumping. Comparing Figs.~\ref{fig:initial44x03}(c1) and (c2), we see that the density interaction ($U>0$, orange) significantly enhances $\bar P_\text{cage}$, particularly when both bosons are already initially inside the cage [Figs.\ref{fig:initial44x03}(c2)]. Saliently, however, the interaction $U$ has a negligible effect if one boson is initially inside the cage and the other outside such that they interact minimally, as in the initial state $(2,4)$ for $x_0=3$ [Figs.\ref{fig:initial44x03}(c3)].

\section{Discussion and generalizations}

Although we have explicitly considered only two particles, the caging mechanism that leads to ultra-strong boundary clustering holds for rather general density interactions with arbitrary numbers of bosons. With $N$ bosons, each Fock state is indexed by a lattice point $\vec x=(x_1,...,x_N)^T$ living in the configuration space $\mathbb{R}^N$~\cite{mandel1995optical,scully1997quantum,bruus2004many,altland2010condensed}, and the density takes distinct integer values in the various hyperplanes, depending on the bosonic occupancy. Given a generic interaction that is a multinomial in the densities $n_1,n_2,...$ at various sites, a monomial $Un_{x_1}n_{x_2}...n_{x_M}$ gives different energy offset in the corresponding hyperplanes, such as to form the barriers of a ``hyper-cage''. For instance, with 4 bosons, an interaction $Un_{x_0}n_{x'_0}$ gives an energy $U$ in the planes where two bosons at fixed at $x_0$ and $x'_0$, $2U$ along lines where three bosons are fixed at either $x_0$ or $x'_0$, etc. Due to the non-locality of the NHSE~\cite{li2021impurity,shen2022non,helbig2020generalized,pan2020non}, caging can occur as long as translation invariance is broken in the configuration space, even if the interaction-induced barriers do not completely enclose any region.

We have revealed how the interplay between the NHSE, topology, and boson-boson interactions can unexpectedly lead to ultra-strong bosonic clustering. This particle clustering strongly exceeds that of NHSE-pumped free bosons, being also facilitated by the interaction-induced hybridization of topological and bulk states and an emergent non-local caging mechanism. Our findings generalize to higher numbers of bosons as well as generic density interactions and topologies, bringing forth a new approach to trapping and controlling bosons. 
Experimentally, demonstrations of non-Hermitian systems have been maturely built upon mechanical arrays~\cite{ghatak2020observation,wang2022non}, photonic~\cite{weidemann2020topological,liang2022dynamic}, electrical circuit~\cite{helbig2020generalized,stegmaier2021topological} and quantum circuit platforms~\cite{shen2023observation}. Various platforms can potentially implement the setup for locally interacting bosons to incorporate the $n_x^2$ term, including optical lattices~\cite{kaufman2015entangling,liang2018observation}, photonic systems~\cite{wang2023experimental,xiao2022bound}, circuit QED~\cite{houck2012chip,devoret2013superconducting}, various quantum simulators~\cite{koh2022simulation,zhang2023many,shen2023proposal,koh2023measurement} as well as photonic resonator arrays, as further elaborated in Appendix.

\appendix

\section{Single-boson and two-boson spectra}\label{app:spectra}

  In the main text, we have computed for our Hamiltonian $H$ [Eq.~1 in the main text] for two bosons at strong non-Hermitian hopping asymmetry $r=4$. The emergence of five distinct bands in the two-boson spectrum, compared to three bands typically seen in single-boson cases, requires careful explanation. Below, we elucidate the existence of these five bands by analyzing the system in the non-interacting limit ($U\to 0$).
  
\begin{figure}[h]
    \subfigure[]{\includegraphics[width=0.15\textwidth]{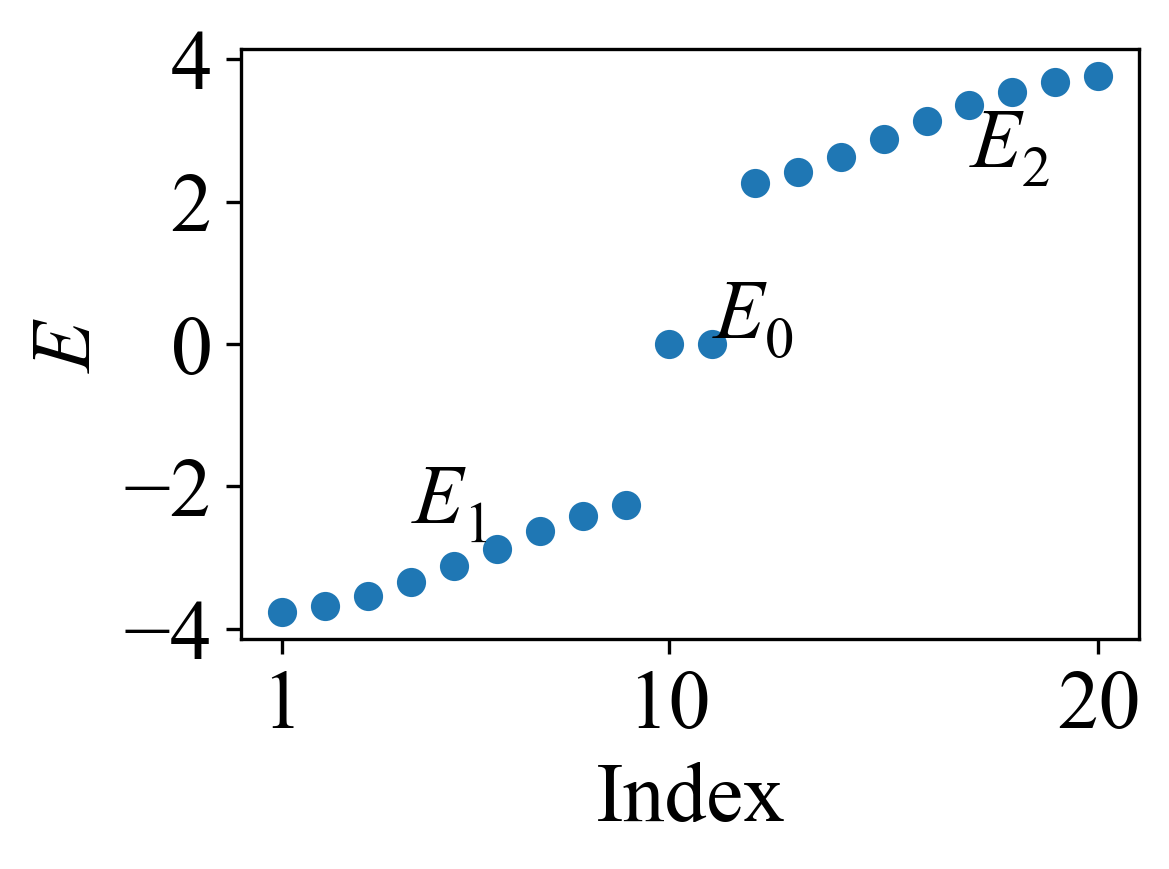}}
    \subfigure[]{\includegraphics[width=0.15\textwidth]{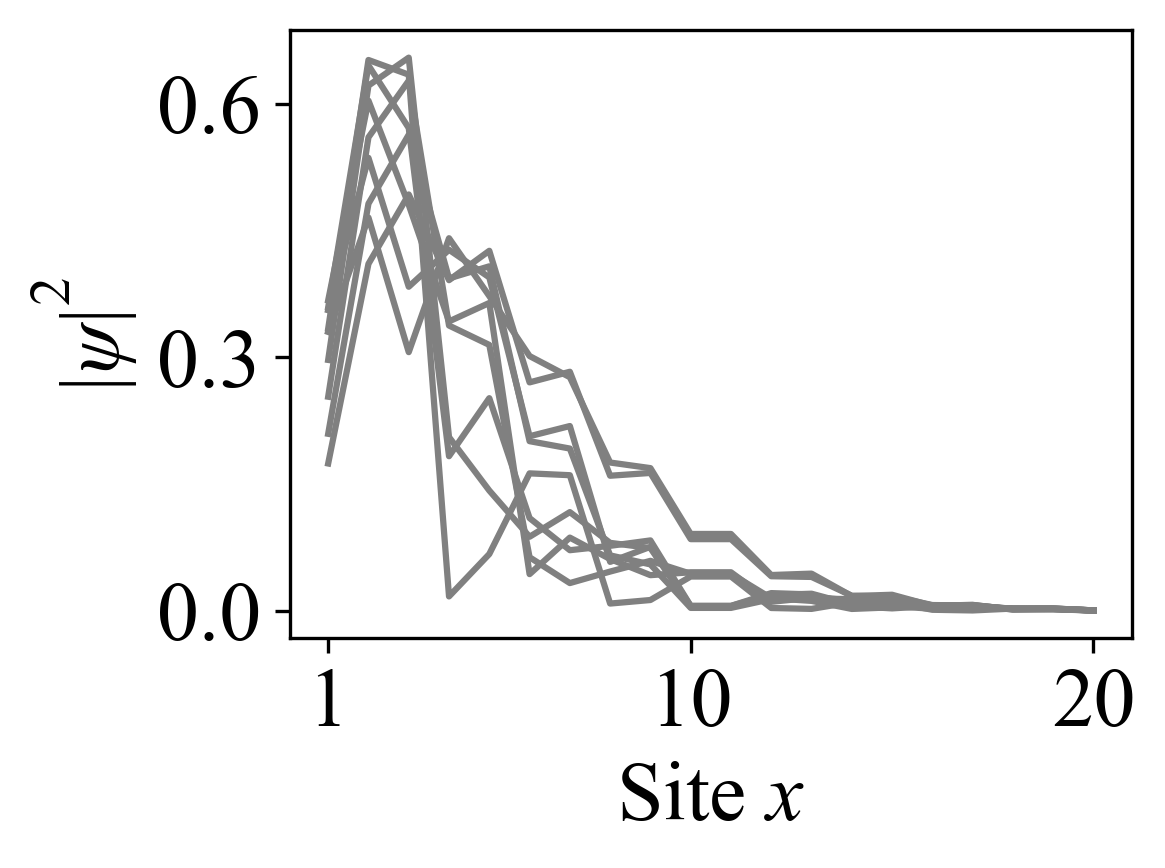}}
    \subfigure[]{\includegraphics[width=0.15\textwidth]{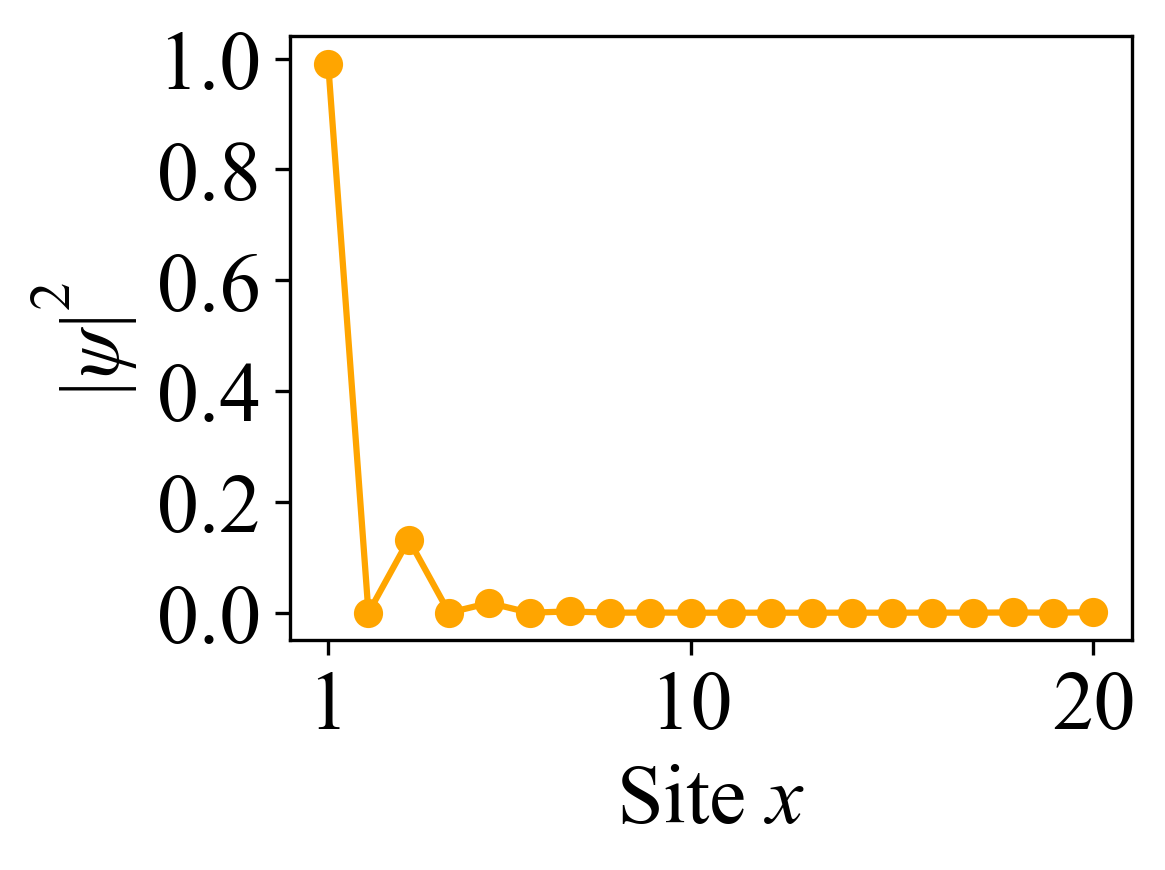}}
      \caption{Single-boson spectrum of the non-Hermitian SSH model [Eq.~\ref{smeq:H} with $U=0$] under open boundary conditions with $t_L=1.6$, $t_R=0.4$ and $t_0=3$. (a) The energy spectrum shows two bulk bands ($E_1$ and $E_2$) separated by topological zero modes ($E_0$). (b) Localization profile $|\psi|^2$ of left boundary-localized states in the bulk bands. (c) Localization of the topological zero mode, is also confined at the left boundary. }
      \label{smfig:sbspectrum}
    \end{figure}

  The single-boson band structure, shown in Fig.~\ref{smfig:sbspectrum}(a), contains two symmetrically gapped bulk bands ($E_1$ and $E_2$) separated by in-gap topological zero modes ($E_0$) at zero energy. The corresponding eigenstates demonstrate distinct localization profiles, with {boundary skin}-localized states for bulk bands [Fig.~\ref{smfig:sbspectrum}(b)] as well as topological zero modes [Fig.~\ref{smfig:sbspectrum}(c)].
  
    \begin{figure}[h]
      \subfigure[]{\includegraphics[width=0.15\textwidth]{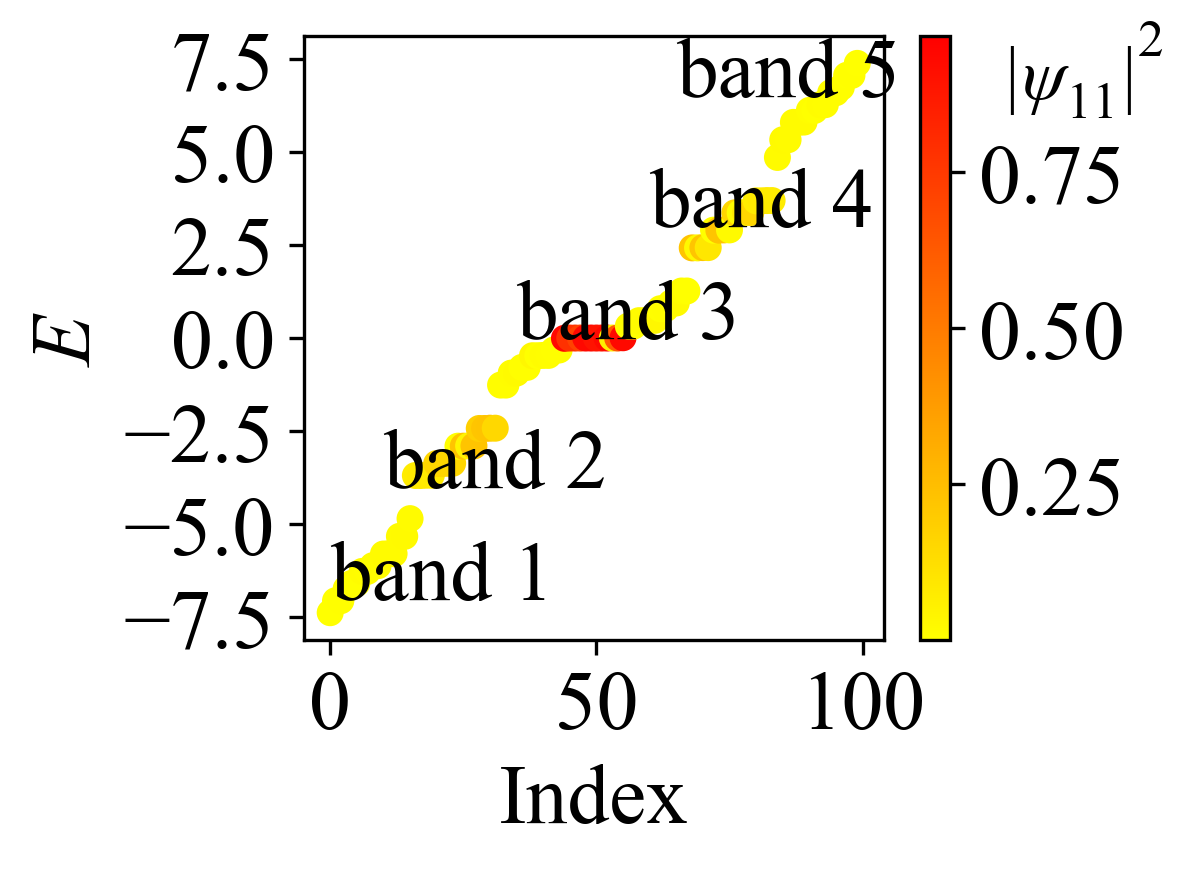}}
      \subfigure[]{\includegraphics[width=0.15\textwidth]{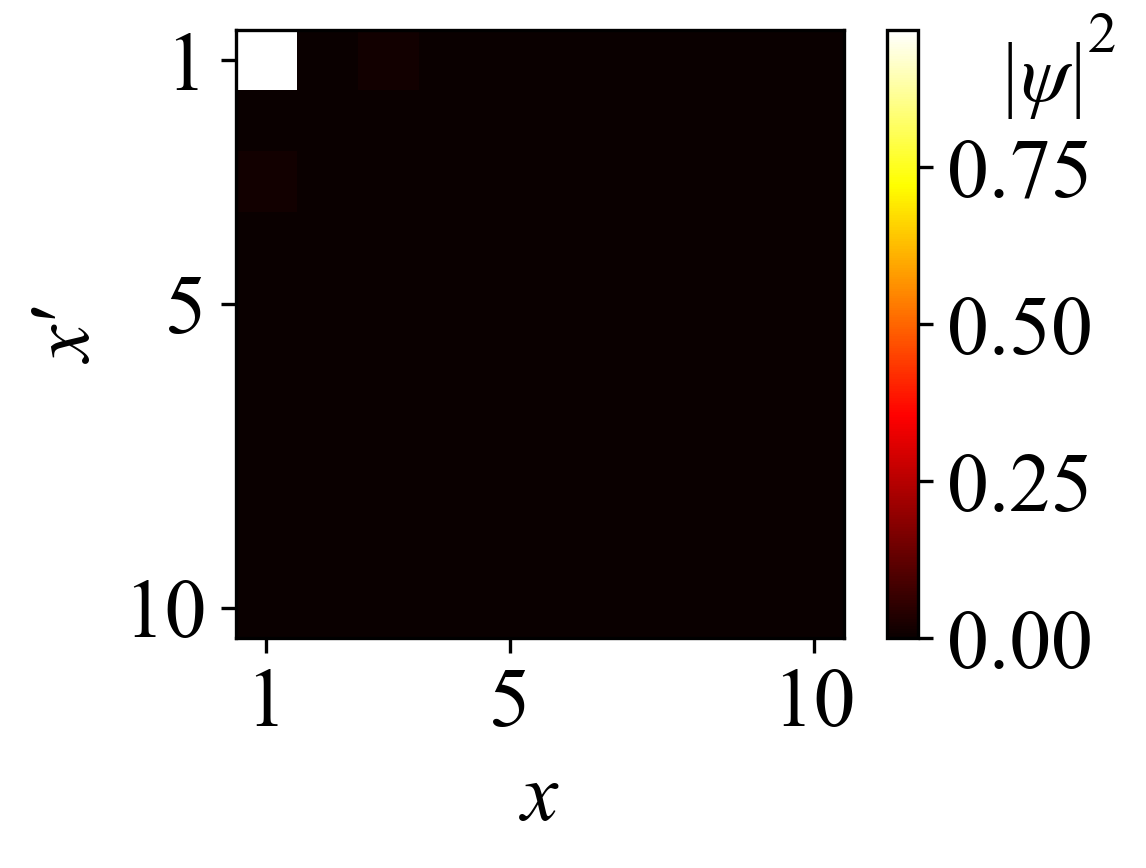}}
      \subfigure[]{\includegraphics[width=0.15\textwidth]{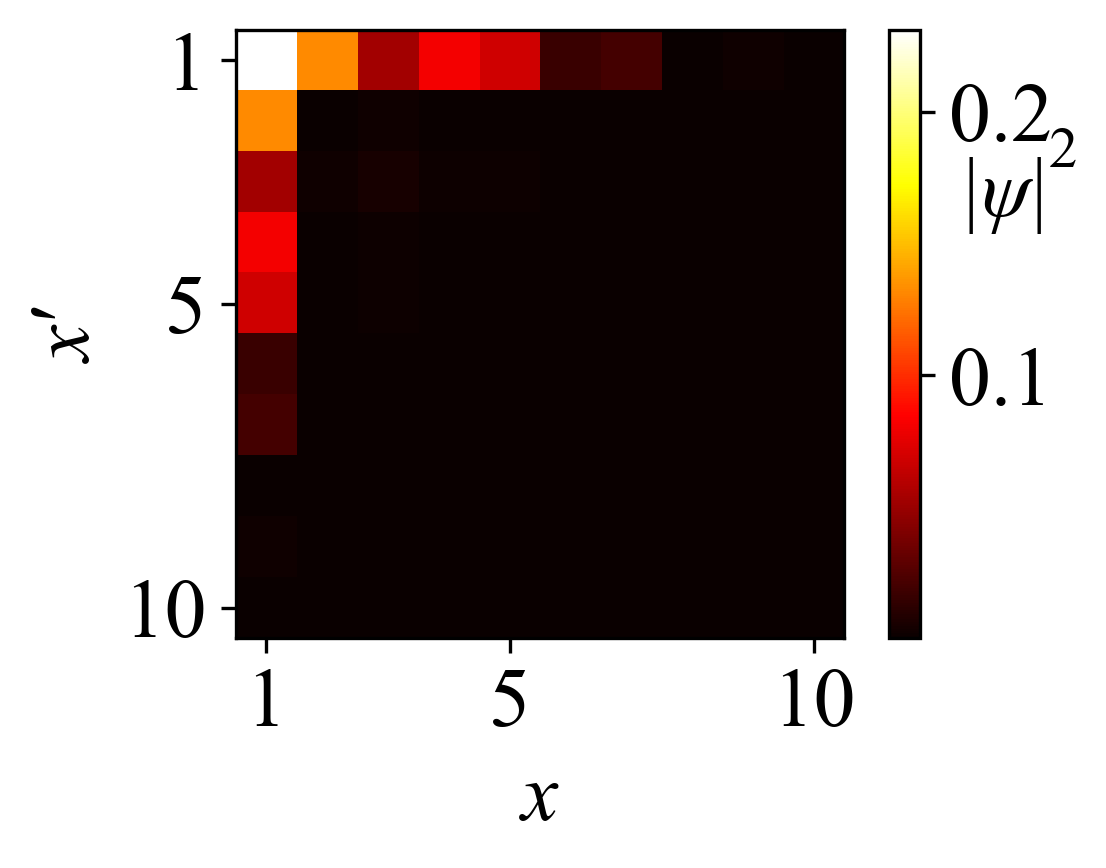}}
        \caption{Two-boson spectrum in the non-interacting limit ($U=0$). (a) Energy spectrum showing five bands arising from combinations of single-particle energies. Band 1 results from hybridization of $E_1 \oplus E_1$. Band 2 results from hybridization of $E_1 (E_0) \oplus E_0 (E_1)$. Band 3 results from the hybridization of $E_1 \oplus E_2$ and $E_0 \oplus E_0$ configurations. Band 4 results from hybridization of $E_2 (E_0) \oplus E_0 (E_2)$. Band 5 results from hybridization of $E_2 \oplus E_2$. (b) Localization pattern of a zero-corner state in band 3. (c) Localization of mixed bulk-topological states in bands 2 and 4, where either $x$ or $x'$ (both not both) are boundary-localized.} 
        \label{smfig:2bspectrum}
      \end{figure}
  
For the two-boson case in the non-interacting limit, the energy spectrum can be systematically constructed through all possible pairwise combinations of the single-particle energies. As shown in Fig.~\ref{smfig:2bspectrum}(a), this results in five distinct bands rather than three. The lowest [band 1 in Fig.~\ref{smfig:2bspectrum}] and highest [band 5 in Fig.~\ref{smfig:2bspectrum}] bands correspond to configurations where both bosons occupy the lower ($E_1 \oplus E_1$) or upper ($E_2 \oplus E_2$) bulk bands, respectively. The intermediate bands 2 and 4 in Fig.~\ref{smfig:2bspectrum} arise from mixed occupations, where one boson occupies the zero mode while the other occupies either the lower ($E_0 \oplus E_1$) or upper ($E_0 \oplus E_2$) bulk band.

Notably, the central band [band 3 in Fig.~\ref{smfig:2bspectrum}] emerges from the merging of two distinct contributions: one where the bosons are distributed between the lower and upper bulk bands ($E_1 \oplus E_2$), and another (much rarer) instance where both bosons occupy the zero mode ($E_0 \oplus E_0$). This explains why we observe five bands rather than six possible energy band combinations, as the energetic proximity of these configurations leads to their mixing into a single band. 
  
The localization profiles shown in Fig.~\ref{smfig:2bspectrum}(b) and (c) reflect this hybridization, exhibiting characteristics that can be understood as superpositions of the single-particle localization profiles.

  \section{Possible physical realizations}\label{smsec:exp}
  
  In the main text, we have proposed a bosonic interacting model in a non-Hermitian SSH lattice, as shown in Eq.1, also reproduced here as Eq.~\ref{smeq:H}:
  \begin{equation}
    \begin{aligned}
        H=&\sum_{x=1}^L t_L b_{2x-1}^{\dagger} b_{2x}+t_R b_{2x}^{\dagger} b_{2x-1}\\&+\sum_{x=1}^{L-1}t_0\left(b_{2x}^{\dagger} b_{2x+1}+b_{2x+1}^{\dagger} b_{2x}\right)+\frac{U}{2} n_{x_0}^2.
    \end{aligned} \label{smeq:H}
    \end{equation}

Below, we discuss how the bosonic interacting model of Eq.~\ref{smeq:H} may be potentially realized using photonic resonator arrays. Arrays of micro-resonators have been extensively utilized to implement one-dimensional (1D) and two-dimensional (2D) photonic lattices, particularly in the investigation of non-Hermitian photonics over the past decade~\cite{chang2014parity,zhao2018topological,zhang2020synthetic}. These designs predominantly employ ring resonators, which can achieve exceptionally high Q factors~\cite{zhao2018topological}.

In Ref.~\cite{zhu2020photonic}, the authors proposed how a non-Hermitian SSH model can be realized using photonic coupled resonant arrays. The non-Hermitian asymmetric coupling can be realized through the judicious incorporation of optical gain and loss elements into unidirectional coupling link rings.

To implement the nonlinear density term, we can employ the Kerr effect which introduces third-order nonlinearity $\chi^{(3)}$. This nonlinear optical effect causes the refractive index of a medium to change with the intensity of light. The third-order nonlinearity can be achieved through natural materials~\cite{shen1984principles,lu2013third} (although the $\chi^{(3)}$ value is relatively small) or by employing cavity QED based on electromagnetically induced transparency (EIT) to achieve an equivalent $\chi^{(3)}$. The latter involves introducing atoms~\cite{chang2014quantum,strater2016floquet} or quantum dots into a cavity~\cite{birnbaum2005photon,kubanek2008two,koch2011three,volz2012ultrafast} and, through appropriate external field control, inducing a target optical field with strong effective nonlinearity.

\section{Bosonic clustering in the Hatano-Nelson model}

\begin{figure*}
  \subfigure[]{\includegraphics[width=0.23\textwidth]{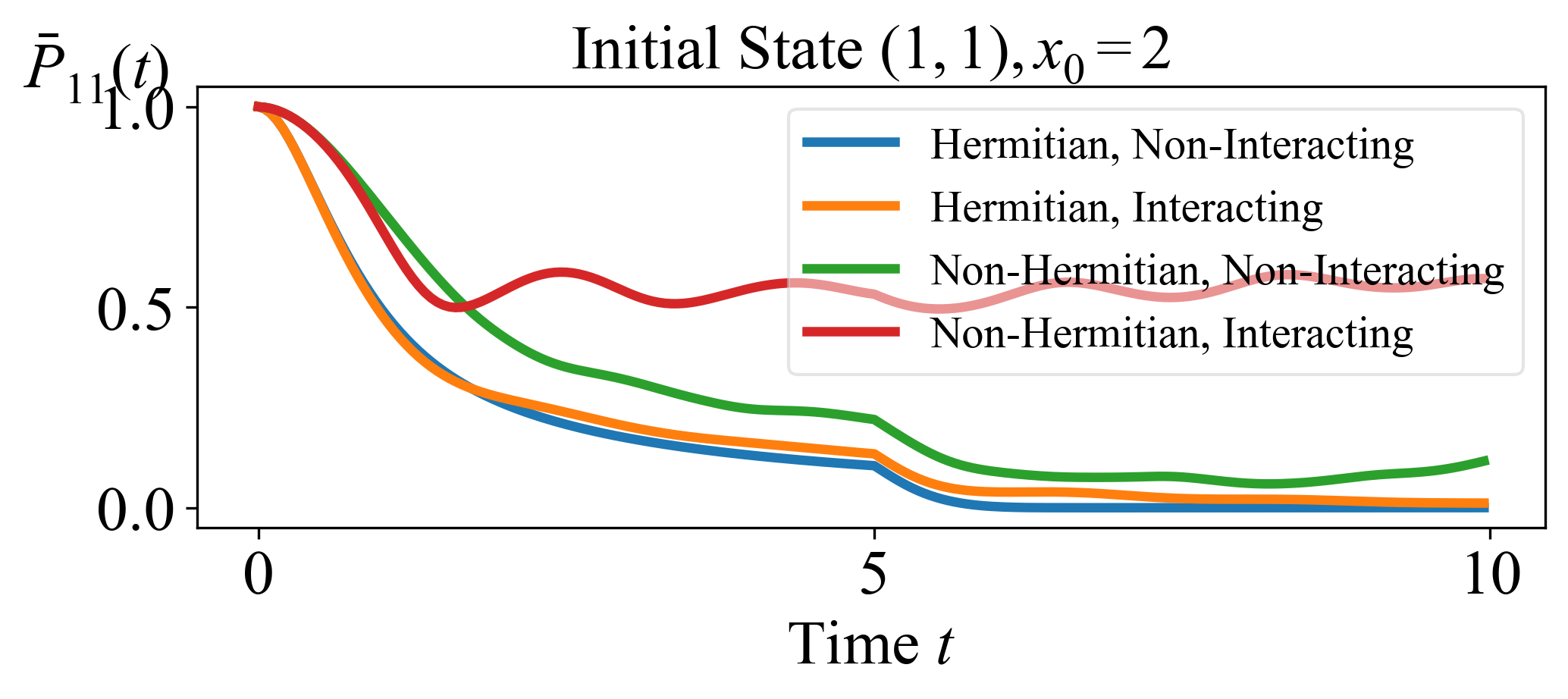}}
  \subfigure[]{\includegraphics[width=0.23\textwidth]{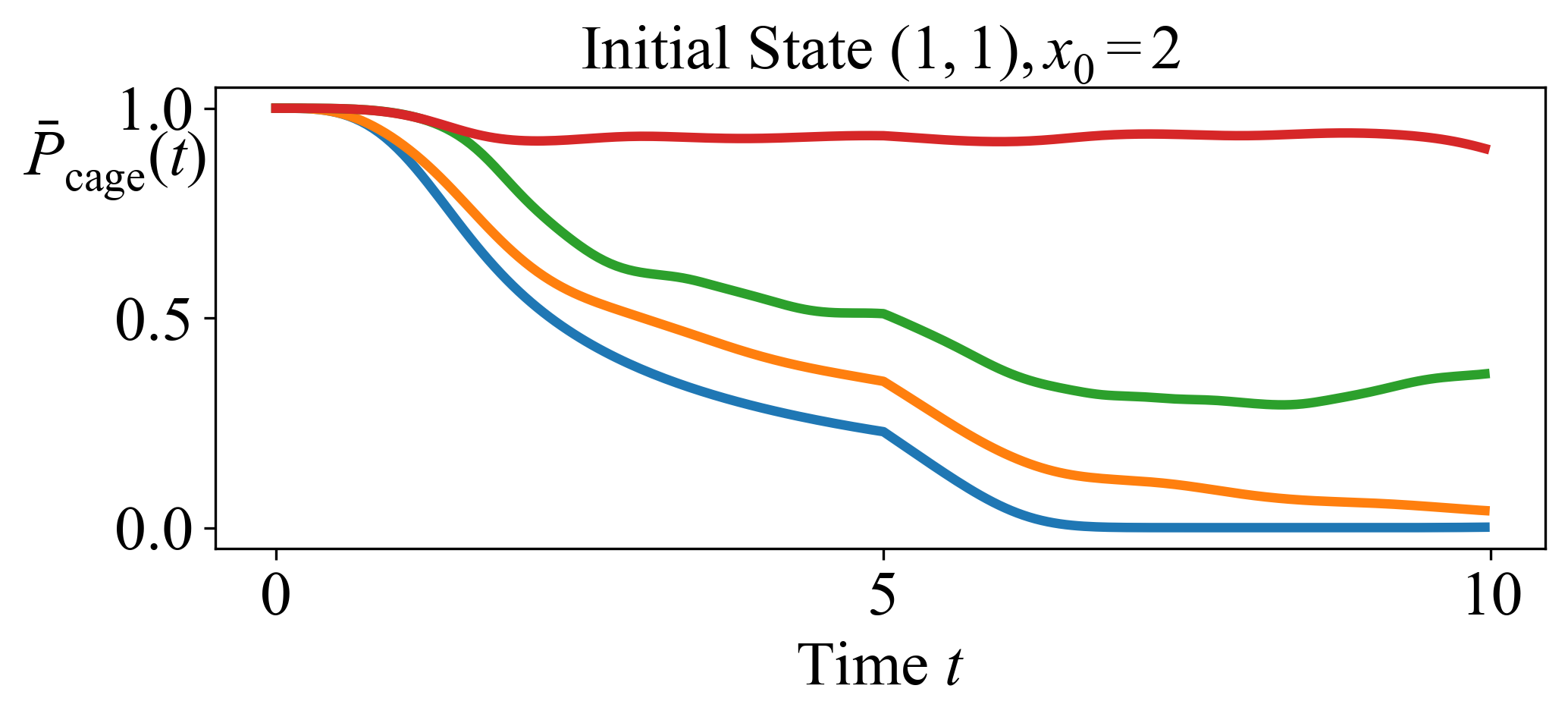}}
  \subfigure[]{\includegraphics[width=0.23\textwidth]{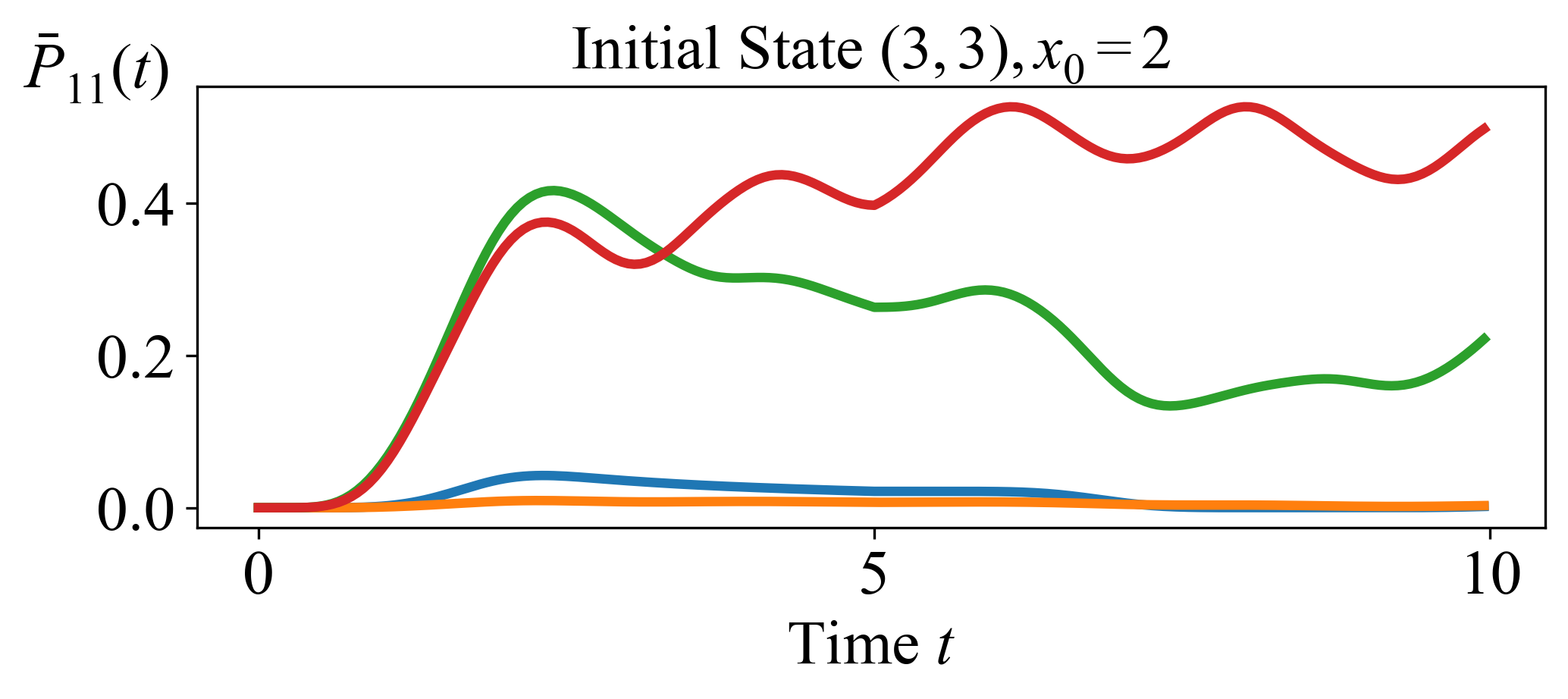}}
  \subfigure[]{\includegraphics[width=0.23\textwidth]{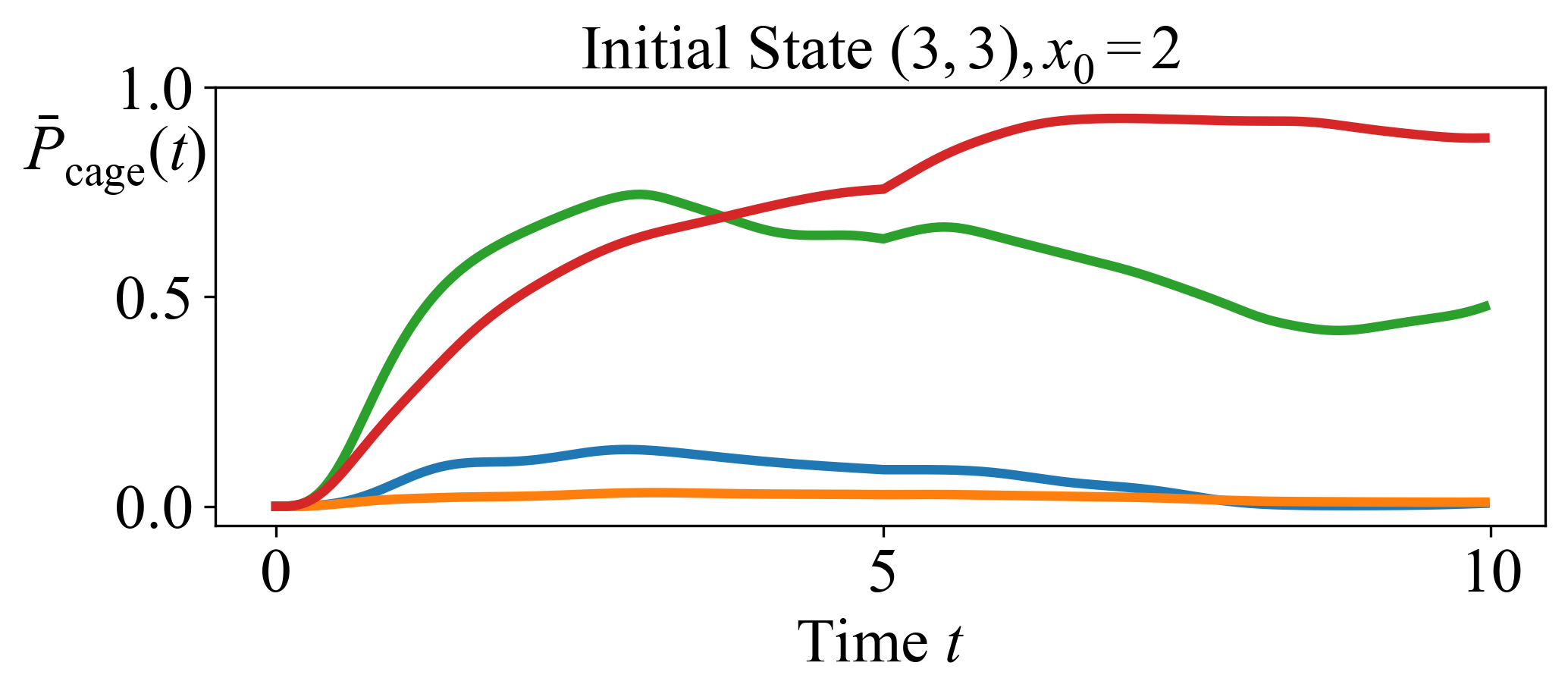}}
  \caption{(a) and (b) show the smoothed two-boson correlation probability $P_{11}(t)$ in Eq.~6 in the main text and the smoothed two-boson probability $\bar P_\text{cage}(t)$ within the $x_0\times x_0$ cage for the initial state $(1,1)$ [inside the cage] at interaction position $x_0=2$. (c) and (d) show the same quantities for the initial state $(3,3)$ [outside the cage] and $x_0=2$. $r=t_L/t_R=4, t_L=1/t_R=2, U=6, L=10$ for the Hatano-Nelson model in Eq.~\ref{smeq:HN}.}
  \label{smfig:HN_cage}
  \end{figure*}

  \begin{figure*}
    \subfigure[]{\includegraphics[width=0.23\textwidth]{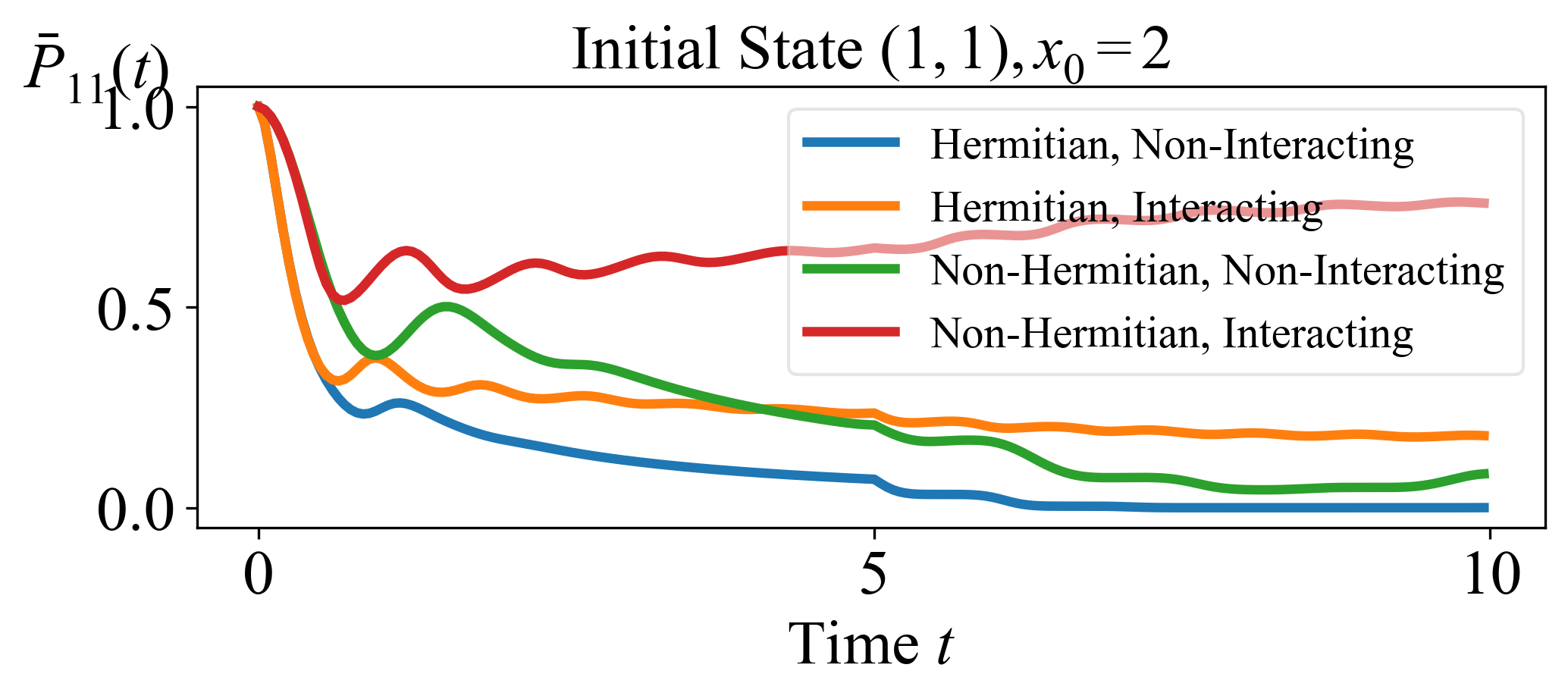}}
    \subfigure[]{\includegraphics[width=0.23\textwidth]{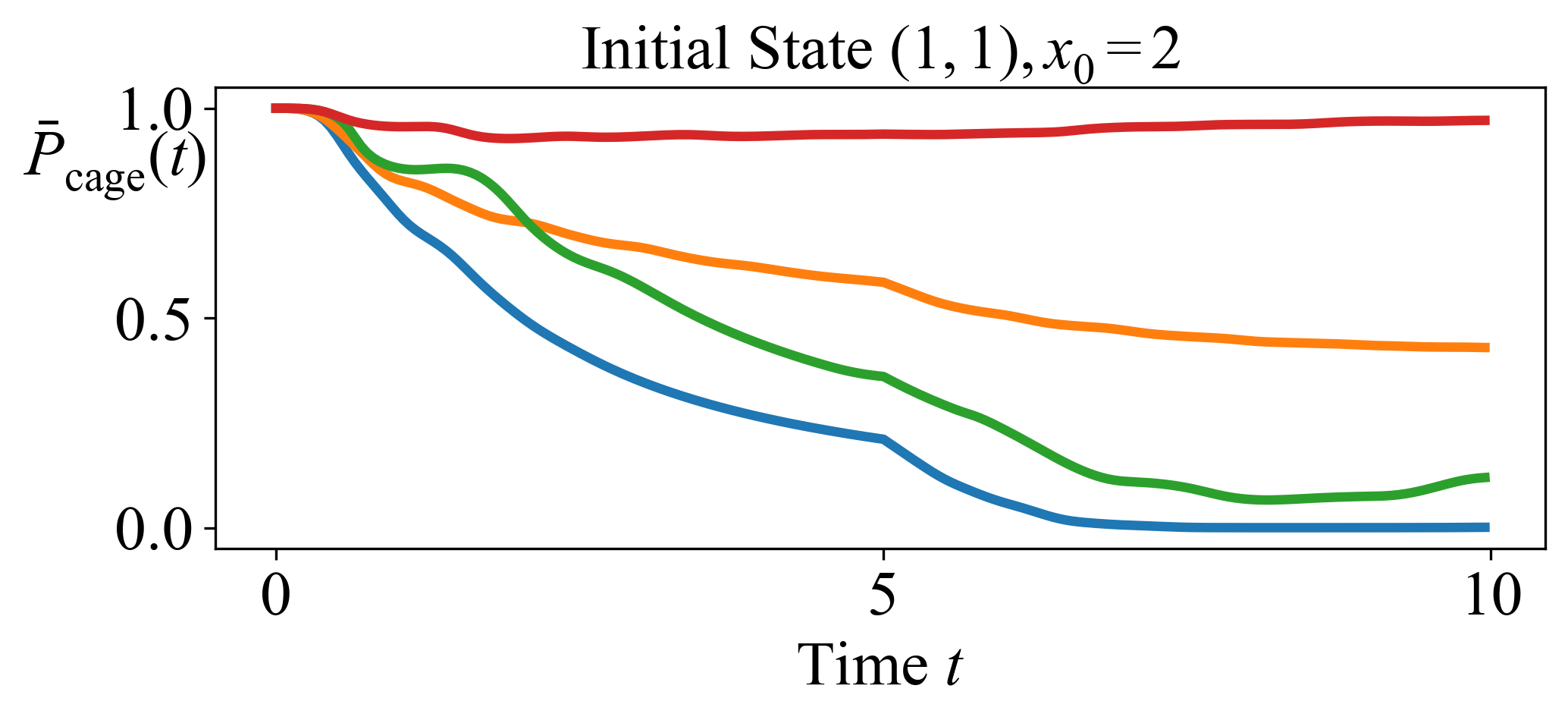}}
    \subfigure[]{\includegraphics[width=0.23\textwidth]{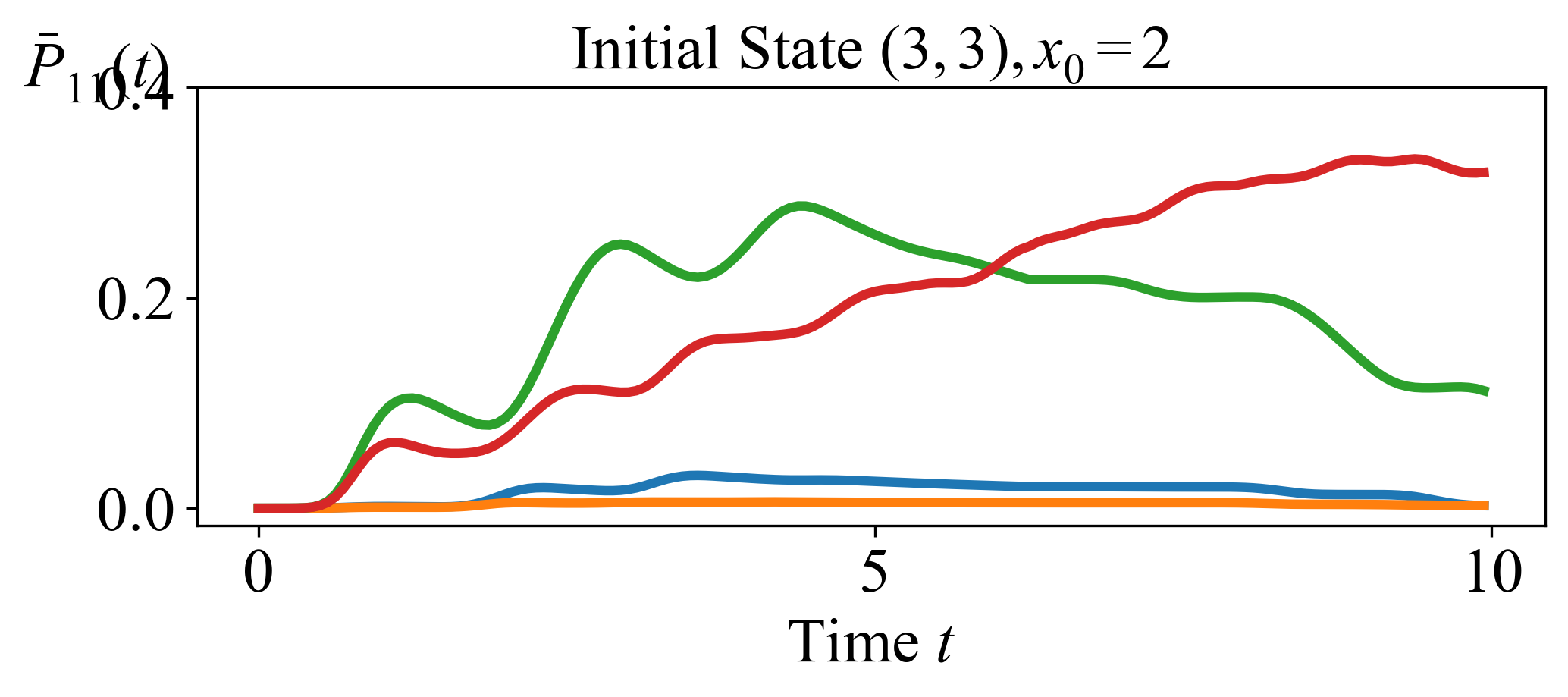}}
    \subfigure[]{\includegraphics[width=0.23\textwidth]{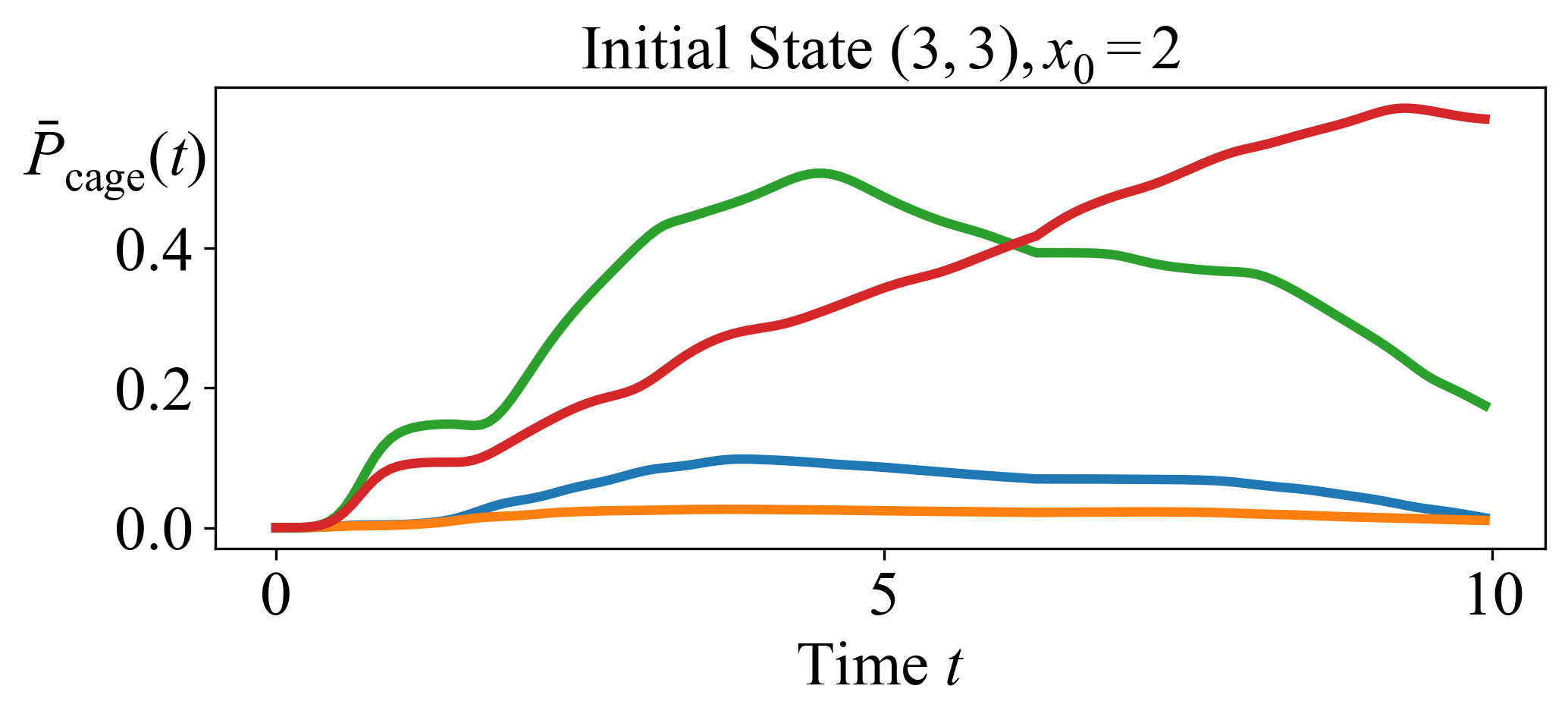}}
    \caption{(a) and (b) show the smoothed two-boson correlation probability $P_{11}(t)$ in Eq.~6 in the main text and the smoothed two-boson probability $\bar P_\text{cage}(t)$ within the $x_0\times x_0$ cage for the initial state $(1,1)$ [inside the cage] at interaction position $x_0=2$. (c) and (d) show the same quantities for the initial state $(3,3)$ [outside the cage] and $x_0=2$. $r=t_L/t_R=4, t_L=1/t_R=2, U=6, L=10$ for the non-Hermitian SSH model in Eq.~1 in the main text with $t_0=t_0' = 1$, $t_L=t'_L = 4.8$, and $t_R=t'_R = 1.2$.}
    \label{smfig:trivial}
    \end{figure*}

    \begin{figure}
      \subfigure[]{\includegraphics[width=0.23\textwidth]{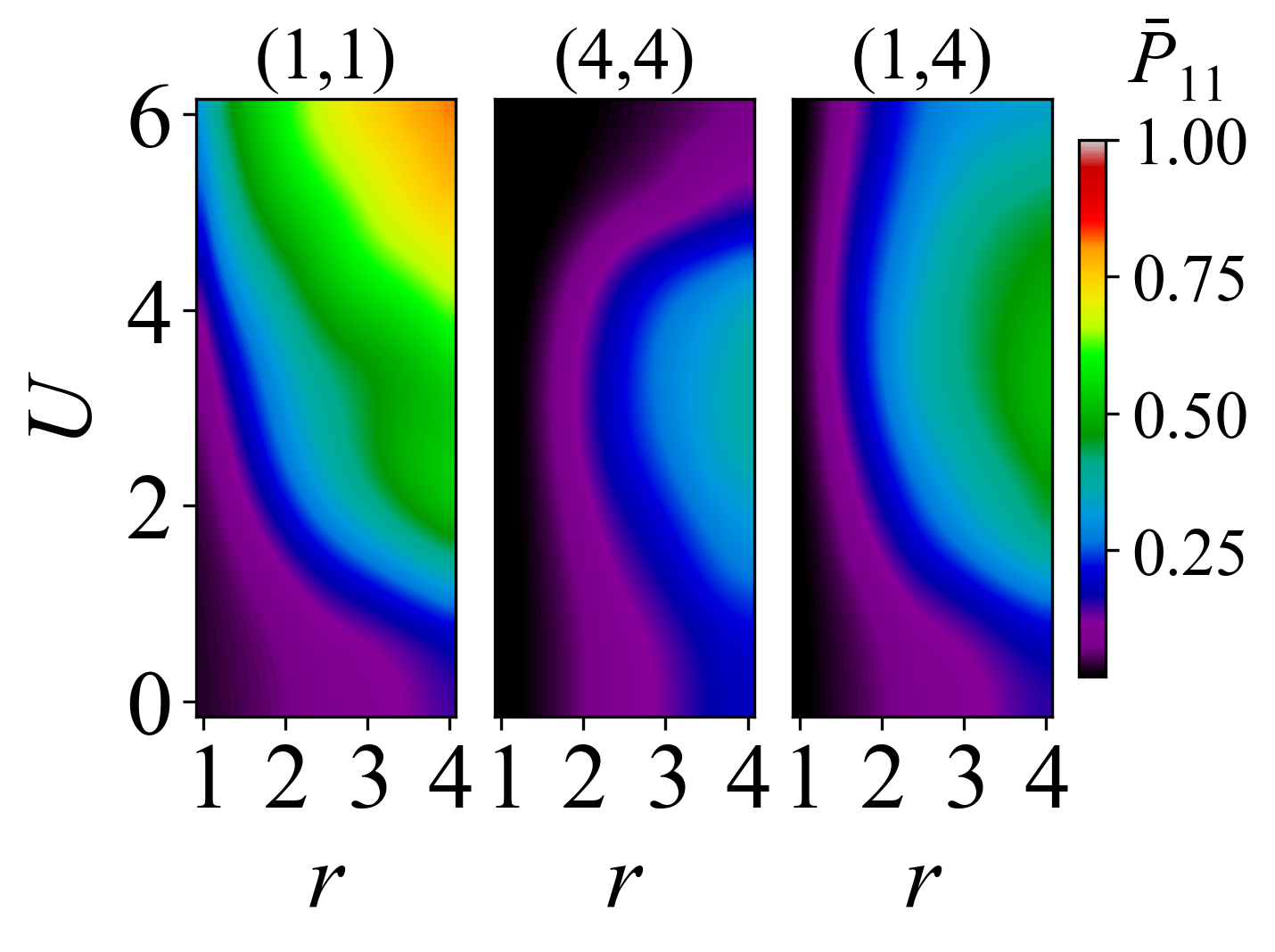}}
      \subfigure[]{\includegraphics[width=0.23\textwidth]{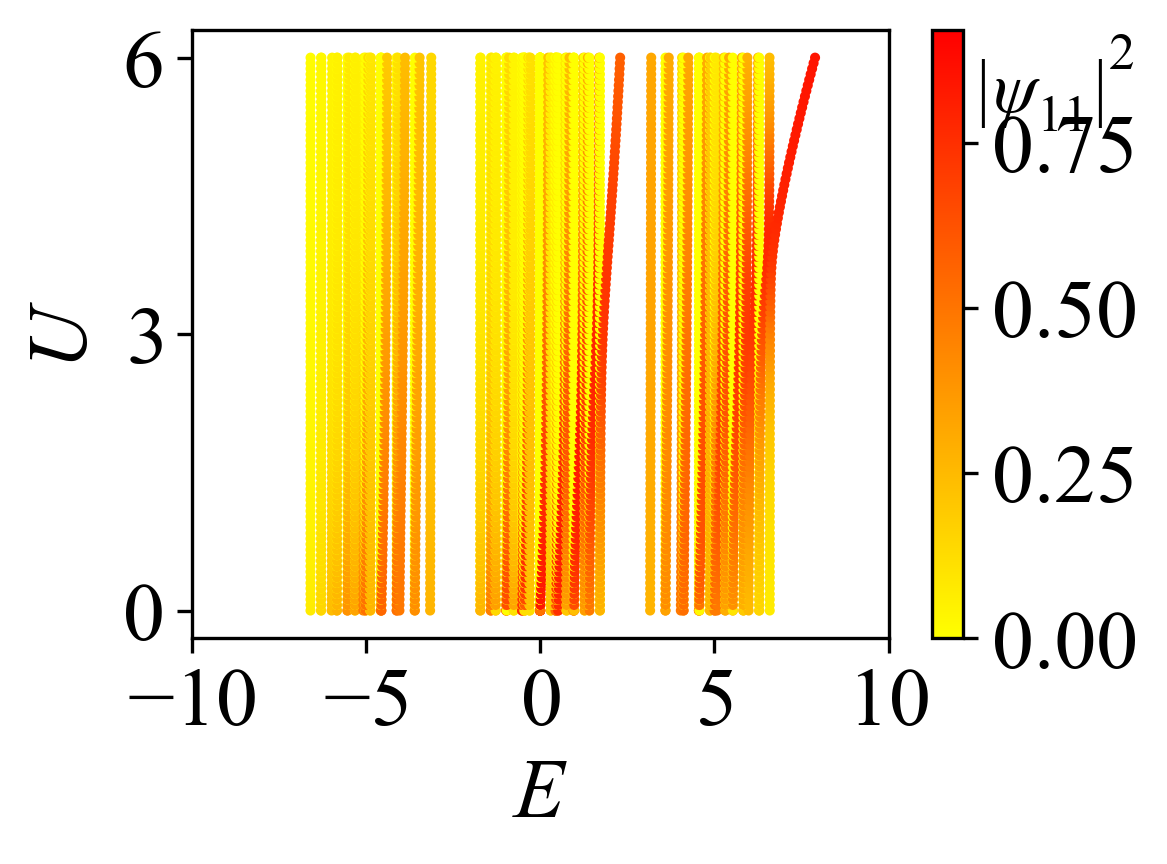}}
      \caption{Extent of two-boson clustering due to a boundary density interaction at $x_0=1$ [Eq.~1 in the main text], and its correspondence with the boundary clustering of 2-boson spectral bands. (a) Time-averaged two-boson clustering probability $\bar{P}_{11}$ [Eq.~6 in the main text] at site $x=1$, in the parameter space of non-Hermitian hopping asymmetry $r$ and density interaction strength $U$. (b) Corresponding two-boson spectra at $r=4$, which features 3 bands only. The topological bands are absent, and the bulk states are localized at the edges. The interesting topological origin described in the main text of the ultra-strong clustering in the nontrivial case is absent in the trivial case. Parameters are $t_0' = 1$, $t'_L = 4.8$, and $t'_R = 1.2$. }
      \label{smfig:trivial_spectrum}
      \end{figure}

In the main text, we have discussed the emergence of ultra-strong bosonic clustering in the non-Hermitian SSH model. Here, we show that the ultra-clustering phenomenon and the caging mechanism are not exclusive to the non-Hermitian SSH model but can also be observed in the Hatano-Nelson model. The Hatano-Nelson model is known to exhibit a non-Hermitian skin effect, where the bulk states are localized at the edges of the system~\cite{hatano1996localization}. After including the bosonic interactions, the interacting Hatano-Nelson Hamiltonian is given by 
\begin{equation}
  H_{\text{HN}}=\sum_{x=1}^{L-1} t_L b_{x}^{\dagger} b_{x+1}+ t_R b_{x+1}^{\dagger} b_{x} + \frac{U}{2} n_{x_0}^2,\label{smeq:HN}
\end{equation}
where $t_L$ and $t_R$ are the left and right hopping amplitudes, respectively. And $U$ is the on-site interaction strength. We will use the same measurement as in the main text—specifically, the smoothed two-boson correlation probability, $\bar{P}_{xx'}(t)$, from Eq.~6 in the main text, which represents the trends of the probability of observing bosons at sites $x$ and $x'$ simultaneously at time $t$—to demonstrate ultra-strong bosonic clustering in the Hatano-Nelson model. 

In Fig.~\ref{smfig:HN_cage}, we show the smoothed two-boson correlation probability $\bar{P}_{11}(t)$ and the smoothed two-boson probability $\bar P_\text{cage}(t)$ within the $x_0\times x_0$ cage for the Hatano-Nelson model. We consider two different initial states similar to in the main text: one where both bosons are initially inside the cage $(1,1)$ and another where both bosons are initially outside the cage $(3,3)$. In both cases, we found the predominantly red curves (non-Hermitian, Interacting) exhibit the same ultra-strong bosonic clustering and caging behavior as in the Hatano-Nelson model. This demonstrates that the ultra-clustering phenomenon and the caging mechanism are not exclusive to the non-Hermitian SSH model but can be observed in other non-Hermitian models as well. 

\section{Bosonic clustering in the non-Hermitian SSH model with trivial topology}

In the main text, we discussed the emergence of ultra-strong bosonic clustering in the non-Hermitian SSH model with nontrivial topology, where $t_L \times t_R = 0.4 \times 1.6 < t_0^2 = 3^2$. Here, we demonstrate that both the ultra-clustering phenomenon and the caging mechanism can also be observed in the non-Hermitian SSH model with trivial topology, where $t'_L \times t'_R > t_0'^2$. 

For this section, we use the parameters $t_0' = 1$, $t'_L = 4.8$, and $t'_R = 1.2$. These values are chosen such that $t_0(t_L + t_R) = t_0'(t'_L + t'_R)$ and $r=t'_L/t'_R=4$, ensuring similar scales and hopping asymmetry for both the nontrivial and trivial cases. Here, $t'_L \times t'_R = 4.8 \times 1.2 > t_0'^2 = 1$, corresponding to the trivial case. The same measurement as in the main text—specifically, the smoothed two-boson correlation probability, $\bar{P}_{xx'}(t)$, from Eq.~6 in the main text—is used to demonstrate ultra-strong bosonic clustering in the non-Hermitian SSH model with trivial topology. And $\bar{P}_{\text{cage}}(t)$, the smoothed two-boson probability within the $x_0\times x_0$ cage, is used to illustrate the caging mechanism.

The ultra-strong bosonic clustering and caging behavior are illustrated in Fig.~\ref{smfig:trivial}. We also consider two different initial states: one where both bosons are initially inside the cage $(1,1)$ and another where both bosons are initially outside the cage $(3,3)$. In both scenarios, we found that the predominantly red curves (non-Hermitian, interacting) exhibit the same ultra-strong bosonic clustering and caging behavior as observed in the non-Hermitian SSH model with nontrivial topology. This demonstrates that the ultra-clustering phenomenon and the caging mechanism are not exclusive to the non-Hermitian SSH model with nontrivial topology; rather, they can also be observed in the non-Hermitian SSH model with trivial topology.

The key difference between the trivial and nontrivial models, as discussed in the main text, lies in the absence of a topological origin in the trivial case. In this case, the topological bands, specifically bands 2 and 4, are no longer present, as illustrated in Fig.~\ref{smfig:trivial_spectrum}(b). Instead, the bulk states are localized at the edges, which is also shown in Fig.~\ref{smfig:trivial_spectrum}(b). The ultra-strong clustering described in the main text, which has a topological origin in the nontrivial case, is absent in the trivial case. Furthermore, the ultra-strong clustering shown in Fig.~\ref{smfig:trivial_spectrum}(a) does not correspond to the boundary clustering of the 2-boson spectral bands in the trivial model.

\end{document}